\documentclass[preprint]{epl}
\usepackage{color}

\title{Gauge Invariance and Fractional Statistics}
\author{A. R. P. Lima\thanks{E-mail: \email{aristeu@fisica.ufc.br}}\inst{1}, R. R. Landim\thanks{E-mail: \email{renan@fisica.ufc.br}}\inst{1}}
\institute{
  \inst{1} Departamento de
F\'{\i}sica, Universidade Federal do Cear\'a, Caixa Postal 6030,
60455-900, Fortaleza, Cear\'a, Brazil}
\begin{document}

\maketitle 

\begin{abstract}
We present a new $(2+1)$-dimensional field theory showing exotic statistics and fractional spin. This theory is achieved through a redefinition of the gauge field $A_{\mu}$. New properties are found. Another way to implement the field redefinition is used with the same results obtained.

\end{abstract}

Since their discovery \cite{d.finkelstein-jmp9,jg.williams-jmp11,mgg.laidlaw-prd3,js.dowker-jppag5,ys.wu-prl52,ys.wu-prl53,dc.ravenel-cmp98} anyons have been given a considerable amount of attention. This is partly due to great theoretical interest in their statistics transmutation properties, and partly to the experimental interest in particles that could account for the fractional quantum Hall effect \cite{mj.bowick-npb271} and the behavior of vortices in superfluid helium films \cite{fdm.haldane-prl55}. Conjectures have been made concerning their importance to the high temperature superconductivity \cite{rb.laughlin-s242} and the two dimensional ferromagnets \cite{l.martina-prb48}.

We follow closely an idea first given by Hagen \cite{cr.hagen-prd31} and successfully executed for the first time by Semenoff \cite{gw.semenoff-prl61}. We modify a three dimensional space-time field theory invariant under $U(1)$ by the addition of a Chern-Simons term. The difference is that now we use a non-minimal coupling of the matter fields with the gauge field $A_{\mu}$. The keypoint here is the construction of the covariant derivative.

When this procedure was first proposed, the action was written in the form
\begin{eqnarray}
S & = & S_{matter}+\int \rm{d}^{3}x(A_{\mu}J^{\mu}+\frac{\alpha}{4\pi}\epsilon^{\mu\nu\rho}A_{\mu}\partial_{\nu}A_{\rho}).
\label{hagen}
\end{eqnarray}

When the actual presentation of the model with exotic statistics was carried out, the action was
\begin{eqnarray}
S & = & \int \rm{d}^{3}x\big(D_{\mu}\phi^{*}D^{\mu}\phi -m^{2}\phi^{*}\phi + \frac{\alpha}{4\pi}\epsilon^{\mu\nu\rho}A_{\mu}\partial_{\nu}A_{\rho}\big),
\label{semenoff}
\end{eqnarray}
with $D_{\mu}=\partial_{\mu}+iA_{\mu}$. It is essentially of the form (\ref{hagen}) except for an additional term: $A_{\mu}A^{\mu}$.

Immediately a question arises whether is possible or not to find a field redefinition able not only to maintain the properties of this action but yet describe properly another field theory in the sense that another interaction can be introduced. Searching for an answer to this question, we found the answer to be, indeed, positive. We stress that this is done in a way to modify the covariant derivative but to keep the CS term intact so that only the terms that interact with the scalar field get altered. Theoretically, one might argue that any gauge invariant term can be added. Experimentally it may not be so.
%

%

We decided to use a field redefinition that generates only gauge invariant terms to obtain the couplings that we intend to consider in our search for fractional statistics models. This redefinition connects the Chern-Simons term, which seems to be of vital importance for fractional statistics with the Maxwell-Chern-Simons (MCS) theory and the Self Dual (SD) theory. We have chosen this redefinition because in \cite{ver.lemes-plb418} and in \cite{ver.lemes-prd58} the Chern-Simons term has been shown to generate the MCS model through field redefinition and in \cite{mam.gomes-jpa38,mam.gomes-el70} that the duality between the two models implies that the field redefinition is the same for both the MCS and SD models.
%

{To the actual calculations we need to follow some conventions.
The metric we use is $(+--)$. The inner product of two p-forms $w_{p}$ and $\eta_{p}$ is denoted by $(w_{p},\eta_{p}) = \int w_{p}\wedge\ast\eta_{p}$, where $\wedge$ is the wedge product and $\ast$ is the Hodge star operator.

Our redefinition comes from the $O$ operator such that, if we define $A'\equiv OA$, we have

\begin{equation}
\frac{1}{2}(dA,dA) + \frac{m}{2}(A,\ast dA) = \frac{m}{2}(A',\ast dA'),
\label{marcio}
\end{equation} 
where $m=\alpha/2\pi$.
}{ The operator $O$ is, then, obtained in an analogous manner to the one carried out in \cite{mam.gomes-jpa38}.
}

Using this redefinition we are led to the following gauge field
\begin{equation}
A^{\prime}_{\mu}\rightarrow A_{\mu}+c_{1}\epsilon_{\mu\nu\rho}\partial^{\nu}A^{\rho}+c_{2}[\partial^{2}( A_{\mu})-\partial_{\mu}\partial_{\nu}A^{\nu}]+\cdots,
\label{redef}
\end{equation}
where $c_{1}$ and $c_{2}$ are numerical constants and the omitted terms are simply powers of the second and third terms and, thus, are not expected to add new features to our investigations.

We then proceed to test our theory having the following covariant derivative
\begin{equation}
D_{\mu}=\partial_{\mu}+\imath A_{\mu}+\imath g\epsilon_{\mu\nu\rho}\partial^{\nu}A^{\rho}
\label{derivada1},
\end{equation}
where $g$ is the coupling parameter.
To this end we employ the Dirac constrained systems quantization method \cite{m.henneaux-qgs,dm.gitman-qfc,pam.dirac-lqm} to the lagrangian (\ref{semenoff}) with (\ref{derivada1}).
%
%
%

It turns out that this model does not present fractional statistics. Previous evidence for this result can be found in the fact that the operator $\epsilon_{\mu\nu\rho}\partial^{\nu}$ alters the nature of the gauge field $ A_{\mu}$. Though it obeys gauge symmetry due to anti-symmetry in its indexes, it does not obey parity symmetry because the Levi-Civita tensor is not invariant under $\mathbf{x} \rightarrow -\mathbf{x}$, thus, leading to an alteration of the parity symmetry of the scalar field interaction terms.

This is completely different for the second term, which has the same parity properties as $ A_{\mu}$. We then expect this model to present not only fractional statistics and spin but new properties still hidden by the minimal coupling used so far. These arguments seem to be in conflict with the fact that th CS term is itself a parity breaking term, but one must take into account that this term is to be coupled to the scalar field and hence enters the action in a different way.

Our lagrangian is, thus, given by (\ref{semenoff}) with

\begin{equation}
D_{\mu}=\partial_{\mu}+iA_{\mu} +ig(\partial^{2} A_{\mu}-\partial_{\mu}\partial_{\nu}A^{\nu}).
\label{derivada2}
\end{equation}
%
%

The next step in the Dirac quantization procedure is to find the momenta canonically conjugated to the fields. They are
\begin{equation}
\pi=\frac{\partial L}{\partial(\partial_{0}\phi^{*})}=D_{0}\phi, \qquad
\pi^{*}=\frac{\partial L}{\partial(\partial_{0}\phi)}= \overline{{D_{0}\phi}},  \label{momen1}
\end{equation}
\begin{equation}
\pi_{0}=\frac{\partial L}{\partial(\partial_{0}A_{0})} \approx 0, \qquad \pi_{\imath}=\frac{\partial L}{\partial(\partial_{0}A_{\imath})} =\frac{\alpha}{4\pi}\epsilon_{\imath \jmath}A_{\jmath}. \label{momen2}
\end{equation}
The symbol $ \approx$ identifies a constraint of the system, ie., a relation that can only be implemented after the calculation of all Poisson brackets. These relations eliminate dynamical variables from the system's phase space.

Performing a Legendre's transformation the Hamiltonian is obtained.
\begin{displaymath}
\mathcal {H}  = p\dot{q}- \mathcal{L}
\end{displaymath}
\begin{eqnarray}
{\cal H} & = & \pi^{*}\pi - A_{0}(\frac{\alpha}{2\pi}\epsilon^{\imath \jmath} + J_{0}) + D_{\imath}\phi \overline{D_{\imath}\phi} + m^{2}\phi^{*}\phi +
{}+\imath g (\partial^{2} A_{0} - \partial_{0} \partial_{\nu} A^{\nu}) 
\times \nonumber\\ &&\times (\phi \partial_{0} \phi^{*} - \phi^{*}\partial_{0}\phi) -2g^{2}(\partial^{2} A_{0}-\partial_{0}\partial_{\nu}A^{\nu})\phi^{*}\phi A_{0}
-2g^{2}[(\partial^{2} A_{0}-\partial_{0}\partial_{\nu}A^{\nu})]^{2}\phi^{*}\phi ,
\label{hamiltoniano1}
\end{eqnarray}
where $ J_{0} = i(\phi^{*}\pi-\pi^{*}\phi)$.

It is interesting to note that an appropriate gauge choice, namely
\begin{equation}
A_{0} \approx 0 \textrm{ and } {\bf \partial}\cdot {\bf A} \approx 0,
\label{gauge}
\end{equation} 
is capable of completely removing the last two lines of the Hamiltonian above as well as the interactions contained in the covariant derivative. This strongly indicates that we are in the correct path.

The consistency of the theory \cite{pam.dirac-lqm} requires the constraint above to be conserved in time. We, then, obtain
\begin{equation}
\frac{\alpha}{2\pi}B + J_{0} + terms \approx 0,
\label{bfield}
\end{equation}
where $ B = \epsilon^{\imath \jmath}\partial_{\imath}A_{\jmath}$ is the magnetic field and the omitted terms vanish in the referred gauge choice. We emphasize that this constraint, considered together with gauge invariance, states that magnetic flux and electric charge cannot be found separately being also precisely the one constraint responsible for the appearance of fractional statistics in this model.

The gauge conditions (\ref{gauge}) and the constraints (\ref{momen2}) determine the gauge field $ A_{\mu}$ to be
\begin{equation}
\hat{A}_{\imath}(\mathbf x) = \frac{\epsilon^{\imath \jmath}}{\alpha} \int \rm{d}^{2}y \frac{x_{\jmath}-y_{\jmath}}{(\mathbf x - \mathbf y)^{2}}J_{0}(\mathbf {y}).
\label{asolution}
\end{equation}

The Hamiltonian is given by (\ref{hamiltoniano1}) with the gauge conditions and ${\bf A}$  replaced by $\hat{ \bf A}$ and the scalar fields satisfying the canonical Poisson Bracket relations 
$ \{\phi(\mathbf x),\pi^{*}(\mathbf y)\}=\{\phi^{*}(\mathbf x),\pi(\mathbf y)\}=\delta^2(\mathbf x - \mathbf y)$.

The gauge field can be written as the divergence of a multivalued operator:
$$
\hat{ \mathbf A}(\mathbf x)  =  \overrightarrow{\partial}\omega, 
$$
where
\begin{equation}
 \omega = \frac{1}{\alpha}\int \rm{d}^{2}y \arctan(\mathbf x - \mathbf y)J_{0}(\mathbf y).
\end{equation}

If we try to remove the interaction through a transformation to the fields with graded algebra \cite{gw.semenoff-prl61}, namely
\begin{eqnarray}
\hat{\phi}(\mathbf x) & = & \exp\big[(i/\alpha)\int \rm{d}^{2}y \arctan(\mathbf x - \mathbf y)J_{0}(\mathbf y)\big]\phi(\mathbf x) \nonumber\\
\hat{\pi}(\mathbf x) & = & \exp\big[(i/\alpha)\int \rm{d}^{2}y \arctan(\mathbf x - \mathbf y)J_{0}(\mathbf y)\big]\pi(\mathbf x) \nonumber\\
\hat{\phi}^{*}(\mathbf x) & = & \exp\big[(i/\alpha)\int \rm{d}^{2}y \arctan(\mathbf x - \mathbf y)J_{0}(\mathbf y)\big]\phi^{*}(\mathbf x) \nonumber\\
\hat{\pi}^{*}(\mathbf x) & = & \exp\big[(i/\alpha)\int \rm{d}^{2}y \arctan(\mathbf x - \mathbf y)J_{0}(\mathbf y)\big]\pi^{*}(\mathbf x) \textrm{,}
\label{transcalcampo}
\end{eqnarray}
we do not obtain an interaction-free Hamiltonian, but
\begin{eqnarray}
{\cal H} & = & \pi^{*}\pi + \phi^{*}(\overleftarrow{\mathbf \partial}+ig\partial^{2}\mathbf{A})\cdot(\overrightarrow{\mathbf \partial}-ig\partial^{2}\mathbf{A})\phi + m^{2}\phi^{*}\phi.
\label{hamilnew}
\end{eqnarray}

Notice that, due to non-commutation, the order of the field operators must be specified in a quantum theory.

It remains for us to evaluate the symmetric Energy-Momentum tensor $ T^{\mu \nu}$. It can be done by coupling the fields to gravity and varying the action with respect to $ g^{\mu \nu}$. 

%
%

The symmetric energy-momentum tensor is
\begin{equation}
T^{\mu \nu} = \overline{D_{\mu}{\phi}}D_{\nu}{\phi} + \overline{D_{\nu}{\phi}}D_{\mu}{\phi} - g_{\mu \nu}(\overline{D_{\lambda}{\phi}}D^{\lambda}{\phi}-m^{2}\phi^{*}\phi).
\end{equation}
%
%

>From the Energy-Momentum tensor we can verify the rotation and translation properties of the system. Let us consider translations first. The gauge dependence of $ \phi$ is balanced by the gauge-dependent translation generator so that the whole transformation is gauge invariant $ \delta_{i}\phi = \{\phi(\mathbf x),\int \rm{d}^{2}y T_{0 \imath}(\mathbf y)\}=(\partial_{\imath}-iA_{\imath}-ig\partial^{2}A_{\imath})\phi(\mathbf x)=D_{\imath}\phi(\mathbf x)$. In order to consider rotations we must evaluate the angular momentum operator $ L = \int \rm{d}^{2}x x_{\imath}\epsilon^{\imath \jmath}T_{0 \jmath}(\mathbf x)$. We will see that it generates a gauge invariant rotation on $\phi$. Replacing $  \mathbf A \textrm{ by } \hat{\mathbf A}$ we obtain the physical rotation and translation operators,
\begin{eqnarray}
P_{\imath} & = & \int \rm {d}^2x( \pi^{*} \partial_{\imath} \phi + \partial_{\imath} \phi^{*} \pi) + \frac{1}{\alpha} \int \rm {d}^2x \int \rm {d}^2y J_{0}(\mathbf x)J_{0}(\mathbf y) \epsilon_{\imath \jmath} \frac{x_{\jmath}-y_{\jmath}}{(\mathbf x - \mathbf y)^2} \nonumber\\
& & + g \int \rm {d}^2x J_{0} (\mathbf x)\partial^{2}{A_{\imath}} \nonumber\\
& = & \int \rm {d}^2x(\pi^{*}\partial_{\imath}\phi+\partial_{\imath}\phi^{*}\pi)
\label{momenoper}
\end{eqnarray}

\begin{eqnarray}
L & = & \int \rm{d}^{2}x (\pi^{*}\mathbf x \times \vec{\partial}\phi+\pi^{*}\mathbf x \times\vec{\partial}\phi^{*}\pi)+ \frac{1}{\alpha}\int \rm {d}^2x\int \rm {d}^2yJ_{0}(\mathbf x)J_{0}(\mathbf y)\frac{\mathbf x \cdot(\mathbf x- \mathbf y)}{(\mathbf x - \mathbf y)^2} \nonumber\\
& & + g\int \rm{d}^{2}x x_{\imath}\epsilon^{\imath \jmath}J_{0}(\mathbf x)\partial^{2}{A_{\imath}} \nonumber \\
& = & \int \rm{d}^{2}x (\pi^{*}\mathbf x \times \vec{\partial}\phi+\pi^{*}\mathbf x \times\vec{\partial}\phi^{*}\pi)+\frac{Q^{2}}{2\alpha}+ g\frac{2\pi}{\alpha}\int \rm{d}^{2}x J_{0}^{2}(\mathbf x)
\label{angulmomoper}
\end{eqnarray}

The first two terms on the righthand side of equation (\ref{angulmomoper}) are the usual ones. The second being the arbitrary spin operator which denotes fractional statistics. We concern ourselves here with the last one. If we bear in mind equation (\ref{bfield}), this term can be written as
\begin{equation}
g\frac{2\pi}{\alpha}\int \rm{d}^{2}x J_{0}^{2}(\mathbf x)=g\frac{\alpha}{2\pi}\int \rm{d}^{2}x B^{2}(\mathbf x)=4g\alpha W_{M}
\label{benergy},
\end{equation}
where $W_{M}$ is the magnetic energy in gaussian units.

Evaluating the Poisson Bracket between the angular momentum operator and the field $ \phi$, we see that it generates a physical rotation (made explicit by the first term), an anomalous spin transformation ($ Q$ term) and a third transformation connected to the magnetic field as evaluated in the point in which the transformation is carried out

\begin{equation}
\{\phi(\mathbf x),L\} = { \mathbf x} \times \vec{\partial} \phi(\mathbf x) - \frac{iQ}{{\alpha}}\phi(\mathbf x)+2igB(\mathbf x)\phi(\mathbf x).
\label{comuta}
\end{equation}

Like the $ Q$ term, the third term cannot be removed by a redefinition of the angular momentum operator without breaking the Poincare algebra $\{K_{\imath},K_{\jmath} \} = \epsilon_{\imath\jmath} L$. We emphasize that this term is new and is to be expected for large magnetic fields $B(\mathbf x)$. 

These results can be obtained  by a slightly different method, which has the same covariant deivative as \cite{gw.semenoff-prl61} and, of course, no change in the scalar field coupling.  Consider the lagrangian

\begin{equation}
\mathcal{L} = D_{\mu}\phi^{*}D^{\mu}\phi -m^{2}\phi^{*}\phi + \frac{\alpha}{4\pi}\epsilon^{\mu\nu\rho}A_{\mu}\partial_{\nu}A_{\rho} + \frac{\kappa}{2} A_{\mu}\partial^{2}\epsilon^{\mu\nu\rho}\partial_{\nu}A_{\rho},
\label{lagdual}
\end{equation}

with $D_{\mu}=\partial_{\mu}+iA_{\mu}$. 

Lagrangian (\ref{lagdual}) is obtained by summing all the terms in (\ref{redef}) that obey the parity symmetry. This is done by considering the properties of the inner products of p-forms. Suppose $A'$ contains only the even powers in (\ref{redef}). We write

\begin{equation}
\frac{m}{2}(A',\ast dA') = \frac{m}{2}(A,\ast d {O'}^{2}A) = \frac{m}{2}(A,\ast dA) + \frac{m\kappa}{2}(A,(\ast d)^3 A).
\label{marciomod}
\end{equation}
This summation is possible because ${O'}^{2}$ is composed by two terms only, the second being $\kappa\partial^{2}$.
%

The same momenta (\ref{momen1}) and (\ref{momen2}) are obtained but a crucial difference comes out in the constraint implied by conservation of $A_{0}\approx 0$, namely

\begin{equation}
J_{0} + \epsilon^{\imath\jmath}(\frac{\alpha}{2\pi} +\kappa\partial^{2} )\partial_{\imath}A_{\jmath}=0.
\end{equation}

This can be modified to 

\begin{equation}
J'_{0} + \frac{\alpha}{2\pi}\epsilon^{\imath\jmath}\partial_{\imath}A_{\jmath} = 0,
\end{equation}
with $J'_{0}\equiv\big(1 +\frac{2\pi\kappa}{\alpha}\partial^{2}\big)^{-1}J_{0}$.

It is now an easy task to find $A'_{\imath}$ up to an arbitrary order in $\frac{2\pi\kappa}{\alpha}\partial^{2}$. We do this to first order. Performing the transformations (\ref{transcalcampo}), we find
\begin{eqnarray}
{\cal H} & = & \pi^{*}\pi + \phi^{*}(\overleftarrow{\mathbf \partial}+i\frac{2\pi\kappa}{\alpha}\partial^{2}\mathbf{A})\cdot(\overrightarrow{\mathbf \partial}-i\frac{2\pi\kappa}{\alpha}\partial^{2}\mathbf{A})\phi + m^{2}\phi^{*}\phi.
\label{hamilnew2}
\end{eqnarray}

Comparing equations (\ref{hamilnew}) and (\ref{hamilnew2}), we see that, in fact, the same result is obtained for the hamiltonian in these two theories. We emphasize that the results (\ref{angulmomoper}) and (\ref{momenoper}) also hold for this model with $g$ replaced by $\frac{2\pi\kappa}{\alpha}$.

Unfortunately not the method used in equation (\ref{semenoff}) with (\ref{derivada2}) nor the one used in (\ref{lagdual}) shows any advantage over the other. This fact lies on the necessity to expand the operator $O$ in order to evaluate the $A_{\mu}$ in the first, or in order to evaluate $J'_{0}$ in terms of $J_{0}$ in the second method.

As a conclusion we would like to say we believe to have found in the parity of the gauge field and of the coupling, which are the same in essence, a fundamental requirement for fractional statistics. The coupling and the gauge field have been constructed here with the focus on gauge invariance and parity. This allowed us to obtain the desired results, which are destroyed if parity is broken like in (\ref{derivada1}). Said in another way, when one tries to introduce coupling with scalar and the gauge field that do not obey parity symmetry one looses fractional statistics! The field redefinition we mentioned is, thus, constituted by the parity obeying terms in (\ref{redef}). It leads to the lagrangian (\ref{lagdual}) as a field theory with fractional statistics and with properties of its own. Lagrangian (\ref{semenoff}) with (\ref{derivada2}) is equally important but is completely analogous to (\ref{semenoff}) itself.

\end{document}